\documentclass{article}
\usepackage{natbib}
\setcitestyle{authoryear,open={(},close={)},aysep={,}}
\let\cite\citep
\newcommand{\shortcite}[1]{(\citeyear{#1})}

\usepackage{times}

\usepackage{soul}
\usepackage{url}
\usepackage[hidelinks]{hyperref}
\usepackage[utf8]{inputenc}
\usepackage[small]{caption}
\usepackage{graphicx}
\usepackage{amsmath}
\usepackage{booktabs}
\usepackage{pifont}
\usepackage{latexsym}
\usepackage{amsfonts}
\usepackage{amssymb}

\newcommand{\p}{\ensuremath{{\rm P}}}
\newcommand{\np}{\ensuremath{{\rm NP}}}
\newcommand{\rclass}{\ensuremath{{\rm R}}}
\newcommand{\rp}{\ensuremath{{\rm RP}}}
\newcommand{\rnc}{\ensuremath{{\rm RNC}}}

\newcommand{\elec}{\ensuremath{\cal E}}

\newtheorem{theorem}{Theorem}
\newtheorem{corollary}[theorem]{Corollary}

\newtheorem{lemma}[theorem]{Lemma}
\newcommand\qedblob{\ding{113}}
\def\literalqed{{\ \nolinebreak\hfill\mbox{\qedblob\quad}}}

\newtheorem{observation}[theorem]{Observation}
\newtheorem{example}[theorem]{Example}

\newenvironment{proofs}{\noindent{\bf Proof.}\hspace*{1em}}{\literalqed\smallskip}

\newcommand{\prob}[3]{
\begin{description}
  \item[Name:] #1
  \item[Given:] #2
  \item[Question:] #3
\end{description}}

\newcommand{\nonameprob}[3]{
\begin{description}
  \item[Given:] #1
  \item[Question:] #2
\end{description}}

\newcommand{\exm}{Exact Perfect Matching}

\newcommand{\exbim}{Exact Perfect Bipartite Matching}
\newcommand{\exbimfull}{Exact Perfect Bipartite Matching}
\newcommand{\exbibm}{Exact Perfect Bipartite $b$-Matching for Multigraphs}
\newcommand{\exexbibm}{Exact Red-Blue Perfect Bipartite $b$-Matching for Multigraphs}

\newcommand{\exbioptm}{Exact Perfect [Bipartite] Matching}
\newcommand{\exbioptbm}{Exact Perfect [Bipartite] $b$-Matching for Multigraphs}

\newcommand{\exexbioptbm}{Exact Red-Blue Perfect [Bipartite] $b$-Matching for Multigraphs}
\newcommand{\exexbioptm}{Exact Red-Blue Perfect [Bipartite] Matching}

\newcommand{\m}{Perfect Matching}
\newcommand{\bm}{Perfect $b$-Matching for Multigraphs}
\newcommand{\excs}{Exact Cycle Sum}
\newcommand{\epbm}{Exact Perfect Bipartite Matching}

\newcommand{\exactperfectmatching}{\exm}

\newcommand{\exactrbbm}{Exact Red-Blue Perfect $b$-Matching for Multigraphs}
\newcommand{\naturalnumber}{\ensuremath{{\mathbb{N}}}}

\title{Insight into Voting Problem Complexity Using Randomized Classes}

\author{
Zack Fitzsimmons\\
 Dept.\ of Math.\ and Computer Science\\
 College of the Holy Cross\\
Worcester, MA 01610 \and
  Edith Hemaspaandra\\
  Department of Computer Science\\
  Rochester Institute of Technology \\
  Rochester, NY 14623}%

\date{July 6, 2022} %

\begin{document}

\maketitle

\begin{abstract}
The first step in classifying the complexity of an NP problem is typically showing the problem in P or NP-complete. This has been a successful first step for many problems, including voting problems. However, in this paper we show that this may not always be the best first step. We consider the problem of constructive control by replacing voters (CCRV) introduced by Loreggia et al.~(2015) for the scoring rule First-Last, which is defined by $\langle 1, 0, \dots, 0, -1\rangle$. We show that this problem is equivalent to Exact Perfect Bipartite Matching, and so CCRV for First-Last can be determined in random polynomial time. So on the one hand, if CCRV for First-Last is NP-complete then RP = NP, which is extremely unlikely. On the other hand, showing that CCRV for First-Last is in P would also show that \epbm\ is in P, which would solve a well-studied 40-year-old open problem.

By considering RP as an option we also gain insight into the complexity of CCRV for 2-Approval, ultimately showing it in \p, which settles the
complexity of the sole open problem in the comprehensive table from Erd\'{e}lyi et al.~(2021).
\end{abstract}

\section{Introduction}

Elections are an important tool used to aggregate the preferences of several agents (voters) over a set of choices (candidates) with applications in areas such as  political domains and multiagent systems in artificial intelligence settings. We consider computational problems relating to elections. Specifically the computational complexity of electoral control.

Electoral control models the actions of an agent with control over an election, referred to as the chair, who modifies this structure to ensure a preferred outcome, e.g., by adding voters to the election to ensure a preferred candidate wins. Types of electoral control model many different types of real-world manipulative actions on elections such as get-out-the-vote drives. The study of electoral control was introduced by Bartholdi, Tovey, and Trick~\shortcite{bar-tov-tri:j:control} who studied the computational complexity of different types of control, including control by adding voters, for several voting rules.
This important initial work led to lots of subsequent work on control 
(see, e.g., Faliszewski and Rothe~\shortcite{fal-rot:b:handbook-comsoc-control-and-bribery}).

A natural model of control
introduced by Loreggia et
al.~\shortcite{lor-nar-ros-bre-wal:c:replacing-control} is 
constructive control by replacing voters (CCRV) in which the 
election chair (for example to avoid detection) replaces a set of 
voters from  the election with the same number of unregistered 
voters in order to ensure a preferred candidate wins.
This model has been thoroughly studied and recent work by Erd{\'e}lyi et al.~\shortcite{erd-nev-reg-rot-yan-zor:j:towards-completing} completes many open cases and includes a comprehensive table of known results. The sole open case is CCRV for 2-Approval elections.

When studying an electoral control problem for a given voting rule, the first step is typically to determine if the action is in \p\ or \np-complete. If the problem is \np-complete, the next steps could involve empirical approaches, parameterized complexity, or approximation~(see, e.g., Rothe and Schend~\shortcite{rot-sch:j:typical-case-challenges} and Dorn and Schlotter~\shortcite{dor-sch:b:trends-parameterized}).
However, some problems in \np\ do not seem to easily be classified as being in \p\ or  \np-complete. An example of such a problem is \exactperfectmatching\ (in which we ask if there exists a perfect matching with exactly a certain number of red edges) introduced by Papadimitriou and Yannakakis~\shortcite{pap-yan:j:spanning-tree} who conjectured it to be \np-complete.
Mulmuley, Vazirani, and Vazirani~\shortcite{mul-vaz-vaz:j:matching} later gave a randomized polynomial-time algorithm thus showing the problem easy.

We consider control by adding voters (CCAV) and control by replacing voters (CCRV) as well as their exact variants (CCAV! and CCRV!) for the scoring rules First-Last (defined by $\langle 1, 0, \dots, 0, -1\rangle$) and 2-Approval,
and link their complexity to the complexity of \exactperfectmatching.
Our results are summarized in Table~\ref{tbl:results}. Equivalence in
Table~\ref{tbl:results} means
polynomial-time disjunctive truth-table (dtt) equivalence (a more flexible notion than many-one equivalence; see the Preliminaries for the formal definition)
though many of the individual reductions are proven with less flexible (typically logspace many-one) reductions.
Since \rp\ is closed under dtt reductions, and Exact Perfect Bipartite Matching and \exm\ are in \rp, the problems polynomial-time dtt equivalent to these problems are also in \rp. If one of the equivalent problems is \np-complete, then $\rp = \np$, which is very unlikely. On the other hand, if one of the equivalent problems is in \p, then so is \exbim,
which would solve a 40-year-old open problem.\footnote{One might wonder why we (and others) do not prove such problems complete for a class such as \rp. The answer to that is that semantically defined classes such as \rp\ may not have complete problems. In particular, \rp\ does not robustly (i.e., in every relativized world) possess many-one complete sets~\cite{sip:c:complete-sets} or even Turing-complete sets~\cite{hem-jai-ver:j:up-turing}.
Of course if $\rp = \p$, then \rp\ does have complete sets.}

\begin{table}\small
\centering
\begin{tabular}{l|r|r}
& \multicolumn{1}{c}{First-Last} & \multicolumn{1}{c}{2-Approval}\\ \hline
CCAV & $\p$~\cite{hem-hem-sch:c:dichotomy-one} & $\p$~\cite{lin:c:manip-k-app} \\ \hline
CCAV! & equiv.\ to \exbimfull\ (Section~\ref{s:FLCCAV})& $\p$~(Theorem~\ref{thm:2app-ccav-exact}) \\ \hline
CCRV & equiv.\ to \exbimfull~(Section~\ref{s:FLCCRV}) & $\p$~(Section~\ref{s:2app-ccrv}) \\ \hline
CCRV! & equiv.\ to \exbimfull~(Section~\ref{s:FLCCRV})  & equiv.\ to \exm~(Section~\ref{sec:2app-exact-ccrv}) \\
\end{tabular}
\caption{Summary of our results}
\label{tbl:results}
\end{table}

Our results establish even better upper bounds than \rp, since \exbim\ and \exm\ are not only in %
\rp, but even in \rnc~\cite{mul-vaz-vaz:j:matching},
which is the class of problems
with efficient randomized parallel algorithms.

Our work is the first to show equivalence between voting problems and \exbioptm.
In fact the only work we found that does something related in voting is
Giorgos~\shortcite{gio:t:multiwinner}, which reduces a multiwinner election problem to \exm\ (but does not show equivalence).
Our approach of showing equivalence to
\exbioptm\ may be useful for other voting problems that have thus-far defied classification as being in \p\ or \np-complete.

\section{Preliminaries}

An election $(C,V)$ is a pair comprised of a set of candidates $C$ and a set of voters $V$ where each
voter $v \in V$ has a total-order vote over the set candidates.

A voting rule $\elec$ is a mapping from an election to a set of candidates referred to as the winners.
Our results are for two natural scoring rules, First-Last and 2-Approval. A scoring rule is defined by a family of scoring vectors of the form $\langle \alpha_1, \alpha_2, \dots, \alpha_m \rangle$ (where $m$ is the number of candidates) with $\alpha_i \ge \alpha_{i+1}$ where a candidate ranked $i$th by a voter receives score $\alpha_i$ from that voter. The rule First-Last is described by the general scoring vector $\langle 1, 0, \dots, 0, -1\rangle$ and 2-Approval is described by the general scoring vector $\langle 1, 1, 0, \dots, 0\rangle$.

\subsection{Election Problems}

We examine the complexity of two types of electoral control, namely, control by adding voters and control by replacing voters. Control by adding voters was introduced by Bartholdi, Tovey, and Trick~\shortcite{bar-tov-tri:j:control}, and control by replacing voters was introduced by Loreggia et al.~\shortcite{lor-nar-ros-bre-wal:c:replacing-control}.

\prob{\elec-Constructive Control by Adding Voters (CCAV)}%
{An election $(C,V)$,
a set of unregistered voters $W$, an integer $k \geq 0$, and a preferred candidate $p$.}%
{Does there exist a set $W' \subseteq W$ such that $\|W'\| \le k$
and $p$ is an \elec-winner of the election $(C, V \cup W')$?}

\prob{\elec-Constructive Control by Replacing Voters (CCRV)}%
{An election $(C,V)$,
a set of unregistered voters $W$, an integer $k \geq 0$, and a preferred candidate $p$.}%
{Do there exist sets $V' \subseteq V$ and $W' \subseteq W$, such that $\|V'\| = \|W'\| \le k$ and $p$ is an \elec-winner of the election $(C, (V - V') \cup W')$?}

We also consider the {\em exact} versions of these problems (introduced by Erd{\'e}lyi~\shortcite{erd-nev-reg-rot-yan-zor:j:towards-completing}) where we ask if it is possible to add or replace exactly $k$ voters. %
We refer to these problems as CCAV! and CCRV!, respectively.

\subsection{Exact Perfect Matching}

This paper explores the connection between the abovementioned election problems with the following graph problems introduced by Papadimitriou and Yannakakis~\shortcite{pap-yan:j:spanning-tree}.

\prob{\exbioptm}%
{A [bipartite] graph $G = (V,E)$, a set
$R \subseteq E$ of red edges, and an integer $k \ge 0$.}%
{Does there exist a perfect matching of $G$ with exactly $k$ red edges, i.e., a set of edges $E' \subseteq E$, exactly $k$ of which are red, such that each $v \in V$ is incident with exactly one edge in $E'$?}

In proving our results we use several different intermediate problems and
define these when they are first used.

\subsection{Computational Complexity}

We assume that the reader is familiar with the
classes \p\ and \np, and what it means for a problem to be \np-complete.
\np-completeness is
typically shown using polynomial-time many-one
reductions.
However, there are many other types of reductions.
Many of our results will use logspace many-one reductions (where the reduction is even computable in logspace) as well as disjunctive truth-table
reductions.
A disjunctive truth-table (dtt) reduction is a generalization of the many-one reduction. A dtt reduction from $X$ to $Y$ outputs a list of strings such that $x \in X$ if and only if at least one of the strings in the list is in $Y$.
Two languages $X$ and $Y$ are equivalent with respect to a given type of reduction if $X$ reduces to $Y$ and $Y$ reduces to $X$, and the type of equivalence (e.g., polynomial-time dtt equivalence) is determined by the
most flexible reduction used in establishing the equivalence.

\rp\ (sometimes called \rclass) is the class of languages $L$ for which there exists a
probabilistic polynomial-time algorithm $A$ such that for all $x \in L$, $A$ accepts with probability at least $1/2$, and for all $x \not \in L$, $A$ always rejects~\cite{gil:j:prob-tms}. Note that $A$ can have false negatives, but not false positives, and by iterating we can make the probability of a false negative exponentially small.
\rnc\ is a subset of \rp, and is defined as the class of languages that can be decided by a randomized algorithm in polylogarithmic time on a parallel computer with a polynomial number of processors (see, e.g., Papadimitriou~\shortcite{pap:b:complexity}).
Since \rnc\ is not known to contain \p, it may not
be closed under polynomial-time reductions, but
\rnc\ is closed under logspace dtt reductions.
We note in passing that the ``\p'' upper bounds listed in Table~\ref{tbl:results} could be replaced by $\p \cap \rnc$, since 2-Approval CCAV! logspace many-one reduces to
2-Approval CCRV by padding (after some preprocessing), CCAV always logspace dtt reduces to CCAV!, and we show that CCRV for 2-Approval and First-Last are in $\rnc$.

For a more detailed description of the concepts introduced in this section, see, e.g., Papadimitriou~\shortcite{pap:b:complexity}.

\section{First-Last Elections}

In this section, we will show that CCRV, CCAV!, and CCRV!
for
First-Last are polynomial-time dtt equivalent to \exbim. These are the first voting problems equivalent to \exbim.
This implies that we are very unlikely to be able to classify these control problems as being in \p\ or being \np-complete: If First-Last CCAV!/CCRV!/CCRV is NP-complete, then $\rp = \np$, which is generally believed to be almost as unlikely as $\p = \np$. And if we were to prove CCAV!/CCRV!/CCRV to be in $\p$, then we would also have shown that \exbim\ is in $\p$.
The proof consists of a large number of reductions, some of which are complicated. To make the proof easier to follow, we will first prove the CCAV! case, which shows the main ideas, and then show how to modify this for CCRV! and CCRV.

\subsection{First-Last-CCAV! Is Equivalent to \exbim}
\label{s:FLCCAV}

The equivalence of First-Last-CCAV!
to \exbim\ follows from the following cycle of reductions:
\begin{enumerate}
\item \excs\ logspace many-one reduces to First-Last-CCAV! (Theorem~\ref{t:CtoCCAV}).
\item First-Last-CCAV! logspace many-one reduces to \exbibm\ (Theorem~\ref{t:CCAVtoM}). \label{it:fl-eccav-exbibm}
\item \exbibm\ logspace many-one reduces to \exbim\ (Theorem~\ref{t:XbtoX}). \label{it:exbibm-exbim}
\item \exbim\ polynomial-time dtt reduces to \excs\ \cite[Proposition 1]{pap-yan:j:spanning-tree}.\footnote{Papadimitriou and Yannakakis~\shortcite{pap-yan:j:spanning-tree} only consider ``polynomial equivalence,'' though it is clear from inspection that some of the reductions in their paper are in fact logspace many-one reductions. 
This particular reduction however computes a perfect bipartite matching (which can be done in polynomial time, but it is not known if this can be done in logarithmic space; we mention that recent work shows this problem is in quasi-NC~\cite{fen-gur-thi:j:bipartite-matching}).
Careful inspection also shows that the reduction from Papadimitriou and Yannakakis~\shortcite{pap-yan:j:spanning-tree} is a dtt %
but not a many-one reduction.}
\end{enumerate}
The above also implies that First-Last-CCAV! is in \rnc, since items~\ref{it:fl-eccav-exbibm} and~\ref{it:exbibm-exbim} imply that 
First-Last-CCAV! logspace many-one reduces to \exbim, \rnc\ is closed under logspace dtt reductions (and so certainly under logspace many-one reductions), and \exbim\ is in \rnc~\cite{mul-vaz-vaz:j:matching}.
\excs\ is defined as follows~\cite{pap-yan:j:spanning-tree}.
\prob{\excs}{Digraph $G$ and integer $k \geq 0$.}
{Is there a set of (vertex) disjoint cycles of total length exactly $k$?}

\exbibm\ is defined as follows.
\prob{\exbibm}
{A bipartite multigraph $G = (V,E)$, a capacity function $b: V \rightarrow \naturalnumber$, a set $R \subseteq E$ of red edges, and an integer $k \geq 0$.}
{Does there exist a perfect $b$-matching of $G$ with exactly $k$ red edges, i.e., a set of edges $E' \subseteq E$, exactly $k$ of which are red, such that each $v \in V$ is incident with exactly $b(v)$ edges in $E'$?}

\begin{theorem}
\label{t:CtoCCAV}
\excs\ logspace many-one reduces to First-Last-CCAV!.
\end{theorem}

\begin{proofs}
We will first show that the obvious attempt at a reduction fails.
Let $G$ be a digraph. Let the candidates be $V(G)$ plus preferred candidate $p$. There are no registered voters. For each arc $(a,b)$ in $G$,
we have an unregistered voter $b > \dots > a$ (i.e., ranking $b$ first and $a$ last) and we ask if we can add exactly $k$ voters such that $p$ is a winner (note that in that case all candidates are tied at 0). It is easy to see that a set of disjoint cycles of total length exactly $k$ corresponds to a set of $k$ added voters such that $p$ is a winner. However, the following example shows that the converse does not hold.

\begin{example}\label{ex:cycle}
Consider the digraph consisting of arcs $(a,b)$, $(b,c)$, $(c,a)$, $(a,d)$, $(d,e)$, and $(e,a)$ (that is, two cycles of length 3, intersecting in $a$) and the corresponding set of voters.

{\centering
\includegraphics[scale=.8]{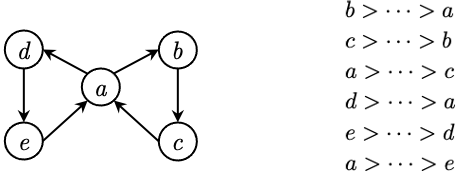}

}

There is no set of disjoint cycles of total length exactly 6 in the given graph. However, if we add all six corresponding voters, $p$ is a winner.
\end{example}

Note that the reduction attempt above {\em will} successfully reduce the version of \excs\ in which the cycles are edge-disjoint rather than (vertex-)disjoint to First-Last-CCAV!. And it is easy to see this reduction is computable in logspace, since each voter corresponds directly to an arc. It remains to show that we can reduce \excs\ to the edge-disjoint version of \excs~(in logspace). This is not hard. Given a digraph $G$, we create a new digraph $G'$ as follows. For every vertex $v$ in $G$, we have an arc $(v,v')$ in $G'$. For every arc $(v,w)$ in $G$, we have an arc $(v',w)$ in $G'$. Set the sum to $2k$.
An example of this construction on the graph from Example~\ref{ex:cycle} is shown below.

{\centering
\includegraphics[scale=0.8]{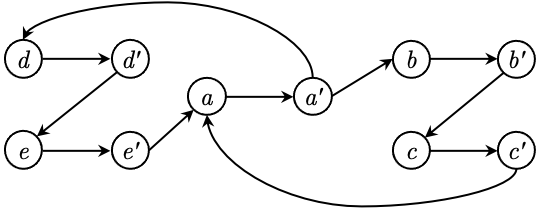}

}

\noindent
 Every cycle that goes through $v$ must go through $(v,v’)$ and so a set of disjoint cycles in $G$ corresponds to a set of edge-disjoint cycles in $G’$ of twice the length.
\end{proofs}

\begin{theorem}
\label{t:CCAVtoM}
First-Last-CCAV! logspace many-one reduces to \exbibm.
\end{theorem}
\begin{proofs}
Let $p$ be the preferred candidate and let $k$ be the number of voters to add.
If there are at least $k$ unregistered voters with $p$ first, we will add only voters that have $p$ first. Then the final score of $p$ will be the score of $p$ from the registered voters plus $k$ and we can easily check if we can add $k$ such voters in such a way that the score of each candidate $a \neq p$ is at most the final score of $p$.

If there are $\ell < k$ unregistered voters with $p$ first, we add all of those (to the registered voters) and subtract $\ell$ from $k$. 
So, all remaining unregistered voters do not have $p$ first. 

Note that if we have not yet determined if control is possible, we are left with an instance in which no unregistered voter has $p$ first (note that in the process, $k$ may have been updated).
If $\ell' < k$ voters do not have $p$ last, add all of those and check if you can add $k - \ell'$ voters that have $p$ last such that $p$ is still a winner.
If there are at least $k$ voters that do not have $p$ last, we will never add a voter that has $p$ last, so delete all unregistered voters that have $p$ last.

Note that after this preprocessing, we have either determined whether control is possible, or we are left with an instance where all unregistered voters give 0 points to $p$.

Let $s_c$ be the score of $c$ from the registered  voters. 
We create the following graph (see Example~\ref{e:CCAVtoM} for an example of the construction).
The vertices are $a, a’$ for every candidate $a \neq p$ and a special vertex $x$. The graph will be bipartite, with the nonprimed vertices (including $x$) in one part and the primed vertices in the other.

For every unregistered voter voting  $b >\dots > a$, we have a red edge $(a,b’)$.
Note that this will give a multigraph, since we can have multiple voters with the same vote.
We also have an ``infinite'' number of nonred edges $(a,a’)$ and an ``infinite'' number of nonred edges $(x,a’)$, where infinite means as many as we could ever need.

Let $M$ be a constant that is large enough so that all $b$-values (to be specified below) are positive. We set the $b$-values as follows: $b(a’) = M$; $b(a) = M + s_a - s_p$; $b(x)$ is set so that the sum of the unprimed $b$-values = the sum of the primed $b$-values, i.e., $b(x) = \sum_{a \neq p} (s_p - s_a)$. If this is negative, then there is no solution.

\begin{example}
\label{e:CCAVtoM}
Consider an instance of First-Last-CCAV!, after the
preprocessing described above has been performed,
with preferred candidate $p$, $k = 2$, registered voters resulting in the following scores: $s_a = 3$, $s_b = -2$, $s_c = -2$, and $s_p = 1$, and five unregistered voters: $b > \dots > a$, $c > \dots > a$, $c > \dots > a$, $b > \dots > c$, and $a > \dots > b$. Below is the corresponding bipartite graph as described in the construction.

{\centering
\includegraphics[scale=0.8]{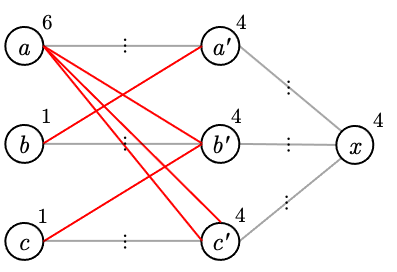}

}

\noindent
$p$ wins after adding the voter voting $b > \dots > a$ and one of the
voters voting $c > \dots > a$. This corresponds to the following perfect matching with exactly two red edges.

{\centering
\includegraphics[scale=0.8]{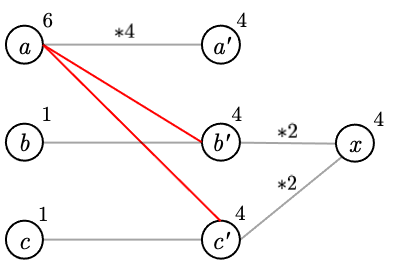}

}

\end{example}

We will show that we can add $k$ unregistered voters such that $p$ becomes a winner if and only if there exists a perfect matching that contains exactly $k$ red edges.

If there is a set of $k$ unregistered voters that we can add so that $p$ becomes a winner, then add the edges corresponding to the added voters to the matching. These are $k$ red edges. We need to show that we can extend this to a perfect matching by adding just nonred edges. 

For $a \neq p$, let $F_a$ be the set of added voters with $a$ first, and let $L_a$ be the set of added voters with $a$ last. Since $p$ is a winner, we know that $s_a + \|F_a\| - \|L_a\| \leq s_p$ ($*$).

The edges corresponding to $L_a$ are incident with $a$, and the edges corresponding to $F_a$ are incident with $a’$. $b(a) = M + s_a - s_p$, and so we need $M + s_a - s_p - \|L_a\|$ nonred edges incident with $a$ to get a perfect matching. All nonred edges incident with $a$ go to $a’$, and there are as many of them as we need. The only thing we need to ensure is that $a’$ can handle that many nonred edges. $b(a’) = M$ and $a’$ is already incident with $\|F_a\|$ red edges. So we need that $M + s_a - s_p - \|L_a\| \leq M - \|F_a\|$. This follows from ($*$).
We then make sure that $a’$ is incident with exactly $b(a’)$ edges by adding as many $(x,a’)$ edges as needed.

For the converse, suppose there is a perfect bipartite $b$-matching
with exactly $k$ red edges. For each red edge $(a,b’)$ add the corresponding voter (who votes $b > \dots > a$). We claim that $p$ is a winner. Consider candidate $a \neq p$. Let $L_a$ be the set of red edges incident with $a$ and let $F_a$ be the set of red edges incident with $a’$. We know that there are $M + s_a - s_p - \|L_a\|$ nonred edges incident with $a$ and that there are $M - \|F_a\|$ nonred edges incident with $a'$. Since all nonred edges incident with $a$ go to $a'$, it follows that $M + s_a - s_p - \|L_a\| \leq M - \|F_a\|$,
which implies that $s_a - s_p + \|F_a\| - \|L_a\| \leq 0$, and so the score of $a$ after addition, $s_a + \|F_a\| - \|L_a\|$, is at most $s_p$.~\end{proofs}

\begin{theorem}
\label{t:XbtoX}
\exbioptbm\ logspace many-one reduces to \exbioptm.
\end{theorem}

The proof of the above theorem generalizes the reduction
from \bm\ to \m\ using the construction from~\cite{tut:j:factor} (see, e.g., Chapter 8 from Berge~\shortcite{ber:b:graphsbook}), and has been deferred
to the appendix.

\def\XbToXproof{%
\begin{proofs}
It is straightforward to reduce \bm\ to \m\ using the construction from~\cite{tut:j:factor} (see, for example,~\citeauthor{ber:b:graphsbook}~(\citeyear{ber:b:graphsbook}, Chapter 8)). 
Given multigraph $G$ and capacity function $b$, replace each vertex $v$ by a complete bipartite graph $(P_v,\{v_1, \ldots, v_{\delta(v)}\})$, where $P_v$ is a set of $\delta(v) - b(v)$ padding vertices (we assume wlog that $\delta(v) \geq b(v)$, otherwise there is no matching). We number the edges incident with $v$ and for each edge $e$ in $G$, if this edge is the $i$th $v$ edge and the $j$th $w$ edge in $G$, we have an edge $(v_i,w_j)$. Call the resulting graph $G'$. Note that $G'$ is bipartite if $G$ is bipartite. To extend the reduction to reduce the exact versions to each other, we color edge $(v_i,w_j)$ red if and only if $e$ is colored red.

It is easy to see that a perfect $b$-matching of $G$ with a $k$ red edges will give a perfect matching of $G'$ with $k$ red edges: Simply take all the edges corresponding to the matching of $G$ and add padding edges to make this a perfect matching. Since padding edges are not red, the obtained perfect matching of $G'$ has exactly $k$ red edges. For the converse, note that any perfect matching of $G'$ consists of padding edges plus a set of edges corresponding to a perfect $b$-matching of $G$. Note that we crucially use that we are looking at {\em perfect} matchings.~\end{proofs}
}

\subsection{First-Last CCRV! and CCRV are Equivalent to \exbim}
\label{s:FLCCRV}

We will now show that CCRV! and CCRV for First-Last are equivalent to \exbim. 
This follows from the following cycle of reductions. 
\begin{enumerate}
\item First-Last-CCAV! logspace many-one reduces to First-Last-CCRV (Theorem~\ref{t:fl-ccav-to-fl-ccrv}).
\item First-Last-CCRV logspace dtt reduces to First-Last-CCRV! (Observation~\ref{obs:ccrv-to-ccrvexact}). \label{it:first-last-ccrv-obs}
\item First-Last-CCRV! logspace many-one reduces to \exbim\ (Theorems~\ref{t:fl-ccrv-to-exexbbm} and~\ref{t:XXbtoX}). \label{it:first-last-ccrv-exbim}
\item \exbim\ polynomial time dtt reduces to First-Last-CCAV! (previous section).
\end{enumerate}
The above also implies that First-Last-CCRV and First-Last-CCRV! are each in \rnc.

We start by noting that control by replacing at most $k$ voters is possible if and only if
for some $\ell$, $0 \leq \ell \leq k$, control by replacing exactly $\ell$ voters is possible.

\begin{observation}\label{obs:ccrv-to-ccrvexact}
For any voting rule $X$, $X$-CCRV logspace dtt reduces to $X$-CCRV!.
\end{observation}

Next, we will show that First-Last-CCAV! reduces to First-Last-CCRV.

\begin{theorem}\label{t:fl-ccav-to-fl-ccrv}
First-Last-CCAV! logspace many-one reduces to First-Last-CCRV (and to First-Last-CCRV!).
\end{theorem}

\begin{proofs}
Consider an instance of First-Last-CCAV! with registered voters
$V$, unregistered voters $W$, preferred candidate $p$, and
limit $k$. As in the proof of Theorem~\ref{t:CCAVtoM} it suffices to consider
First-Last-CCAV! where $p$ receives 0 points from each unregistered voter.

Let $s_p$ denote the score of $p$ from the registered voters. Note that
in a First-Last election the sum of the scores
of all candidates is 0. And so if $s_p$ is negative, 
$p$ cannot be made a winner by exact adding. 

So assume $s_p \geq 0$. We now construct an instance of First-Last-CCRV as follows. Let the candidate set
consist of $C$ with $2k$ additional candidates: 
$a_1, \dots, a_k$ and $b_1, \dots, b_k$. Update the preferences for each voter so that these
candidates receive 0 points from each registered and unregistered voter. Except for adding these
new candidates to the votes, the set of unregistered voters remains the same. However, for each
$i,\ 1 \le i \le k$, add $s_p+1$ voters voting $a_i > \dots > b_i$ to the set of registered voters.
Now each $a_i$ candidate has score $s_p+1$, each $b_i$ candidate has score $-s_p-1$, and the scores
of the remaining candidates are unchanged.

If control is possible by adding exactly $k$ voters such that $p$ wins, then control by replacing voters is possible by replacing $k$ voters, with the $i$th voter voting $a_i > \dots > b_i$ for $1 \leq i \leq k$, with those same $k$ unregistered voters. This
decreases the score of each $a_i$ candidate by 1 so that 
they tie with $p$, and it is easy to see that $p$ wins.

For the converse, suppose that there is a way to replace at most $k$ voters such that $p$ wins. Since the score of each of the $k$ $a_i$
candidates must decrease by 1, and each unregistered voter gives 0 points to $a_i$,
at least one of the voters voting $a_i > \dots > b_i$ must be replaced for each $i$. 
It follows that we replace exactly one voter voting $a_i > \dots > b_i$ for each $i$, and
it is straightforward to see that the $k$ unregistered voters replacing the $a_i$ voters correspond to $k$ voters that can be added to $V$ so that $p$ wins by exact control by adding voters.

The same reduction reduces to CCRV!.~\end{proofs}

And finally, we will show that First-Last-CCRV! reduces to \exbim. Note that we can view CCRV! as CCAV! where we have two sets of unregistered voters. This suggests reducing to the following variation of \exbibm.
\prob{\exexbibm}
{A bipartite multigraph $G = (V,E)$, a capacity function $b: V \rightarrow \naturalnumber$, disjoint sets $R \subseteq E$ of red edges and $B \subseteq E$ of blue edges, an integer $k \geq 0$, and an integer $\ell \geq 0$.}
{Does there exist a perfect $b$-matching of $G$ with exactly $k$ red edges and exactly $\ell$ blue edges, i.e., a set of edges $E' \subseteq E$, exactly $k$ of which are red and exactly $\ell$ of which are blue, such that each $v \in V$ is incident with exactly $b(v)$ edges in $E'$?}

\begin{theorem}\label{t:fl-ccrv-to-exexbbm}
First-Last-CCRV! logspace many-one reduces to \exexbibm.
\end{theorem}

The proof is deferred to the appendix.

\begin{theorem}
\label{t:XXbtoX}
\exexbioptbm\ logspace many-one reduces to \exbioptm.
\end{theorem}

\begin{proofs}
The construction from the proof of Theorem~\ref{t:XbtoX} (when we color edges in the constructed graph with the color of their corresponding edges in the multigraph) reduces \exexbioptbm\ to \exexbioptm.

It remains to reduce \exexbioptm\ to \exbioptm. 

Let $G$ be the graph with red and blue edges, and let $k$ and $\ell$ be the exact numbers of red and blue edges we want in the perfect matching. Without loss of generality, assume that $k, \ell < n$, where $n$ is the number of vertices (note that a perfect matching contains $n/2$ edges). Adapting approaches from Papadimitriou and Yannakakis~(\citeyear{pap-yan:j:spanning-tree}, Proposition 1), we replace each blue edge $(u,v)$ by a path of length $2n-1$ colored red-nonred alternately: %

{\centering
\includegraphics[width=0.4\textwidth]{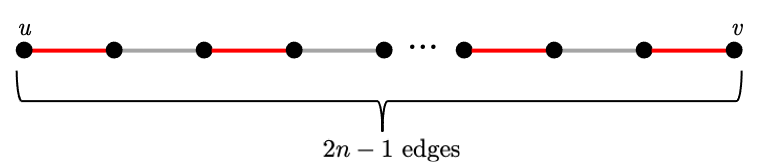}

}

\noindent
Call this graph $G'$. Note that $G'$ is bipartite if $G$ is bipartite and that $G'$ uses only the color red. We claim that $G$ has a perfect matching with $k$ red and $\ell$ blue edges if and only if $G'$ has a perfect matching with $\ell n + k$ red edges. Consider a perfect matching of $G$ with $k$ red and $\ell$ blue edges. We will modify this to obtain a perfect matching for $G'$.
For each blue edge in $G$ consider the length $2n-1$ path in $G'$ that corresponds to this edge. If the blue edge is in the matching of $G$, replace this edge by the $n$ red edges on the path. If the blue edge is not in the matching, put the $n-1$ nonred edges on the path in the matching. This gives a perfect matching of $G'$ with the required number of red edges. For the converse, consider a perfect matching of $G'$ with $\ell n + k$ red edges. Note that every path corresponding to a blue edge in $G$ contributes $n$ or 0 red edges to the perfect matching of $G'$. Also note that the matching of $G'$ contains fewer than $n$ nonpath edges. It follows that exactly $\ell$ of the paths corresponding to a blue edge contribute $n$ red edges to the matching. These are exactly the blue edges in the perfect matching for $G$. And the remaining edges in the matching of $G'$ ($k$ of which are red) are the remaining edges in the perfect matching of $G$.~\end{proofs}

\section{2-Approval Elections}

In the previous section we found that control by replacing voters for First-Last is equivalent to \exbim\ and so it is unlikely to be \np-complete, and showing that it is in \p\ would solve an important open question in matching.
Linking
the complexity of a problem to \exbioptm\ and showing that the problem is in \rp\ can be a useful approach for gaining insight into the complexity of a voting problem which has resisted classification.
In fact we find an $\rp$ (and $\rnc$) upper bound for the complexity of 2-Approval-CCRV by showing that it logspace dtt reduces to \exactperfectmatching\ (Corollary~\ref{cor:2-app-ccrv-exact}). 
Notably,
2-Approval-CCRV is the sole remaining case in the 11-by-12 table
from Erd{\'e}lyi et al.~\shortcite{erd-nev-reg-rot-yan-zor:j:towards-completing}, and 
Corollary~\ref{cor:2-app-ccrv-exact} shows that this problem is easy, since it is in \rp\ and it is widely assumed that $\p = \rp$. This is a significant step in understanding the complexity of this problem. The obvious question is whether we can strengthen this result by showing equivalence with \exactperfectmatching\ or by showing that the problem is in $\p$. In Section~\ref{s:2app-ccrv}, we show the latter, by reducing to a polynomial-time computable matching problem.

\subsection{2-Approval-CCRV! is Equivalent to Exact Perfect Matching}\label{sec:2app-exact-ccrv}

We show that 2-Approval-CCRV! is equivalent to \exactperfectmatching\ through the following cycle of reductions.
\begin{enumerate}
    \item 2-Approval-CCRV! logspace many-one reduces to \exactrbbm\ (Theorem~\ref{thm:2app-ccrv-exact-upper}). \label{it:2app-eccrv-erbbm}
    \item \exactrbbm\ logspace many-one reduces to \exactperfectmatching\ (Theorem~\ref{t:XXbtoX}). \label{it:erbbm-to-exm}
    \item \exactperfectmatching\ logspace many-one reduces to
    2-Approval-CCRV! (Theorem~\ref{thm:match-2-app-exact-ccrv}).
\end{enumerate}
As a result of Observation~\ref{obs:ccrv-to-ccrvexact} and
items~\ref{it:2app-eccrv-erbbm} and~\ref{it:erbbm-to-exm} we have the following corollary.
\begin{corollary}\label{cor:2-app-ccrv-exact}
2-Approval-CCRV logspace dtt reduces to \exactperfectmatching.
\end{corollary}

This implies that 2-Approval-CCRV is in \rnc. Additionally, the above statements imply 2-Approval-CCRV! is in \rnc.

\begin{theorem}\label{thm:2app-ccrv-exact-upper}
2-Approval-CCRV! logspace many-one reduces to \exactrbbm.
\end{theorem}

\begin{proofs}
Consider an instance of 2-Approval-CCRV! with
candidates $C$, registered voters $X$, unregistered voters $Y$, preferred candidate $p$, and number $\ell$.
As also done %
in the proof of Theorem~\ref{t:fl-ccrv-to-exexbbm}, we can
view this problem as a version of exact control by adding
voters where we ask if it is possible to add (to the empty set)
$k = \|X\|-\ell$ voters from $X$ and $\ell$ voters from $Y$ such that $p$ wins. Without loss of generality we assume that $0 \leq k \leq \|X\|$ and $0 \leq \ell \leq \|Y\|$.

Note %
we can assume that we will
add as many voters that approve of $p$ as possible. This fixes the final score of $p$ which we denote by
$s_p = \min(k, \text{\#voters in $X$ approving $p$}) + \min(\ell, \text{\#voters in $Y$ approving $p$})$. 

We will now construct the instance of \exactrbbm.
Let the set of vertices in the constructed graph $V(G) = C \cup \{x\}$, where $x$ is a vertex used to ensure that the resulting matching is perfect. For each $c \in C$ let $b(c) = s_{p}$. %
Let $b(x) = \|C\|s_p - 2(k+\ell)$.
If this is negative, control is not possible (since $2(k+\ell)$ is the total number of points over all candidates).
The set of edges is defined as follows.

\smallskip
\noindent
{\bf Red edges:} For each voter in $X$ that votes $\{a,b\} > \dots$ add a red edge $(a,b)$.

\noindent
{\bf Blue edges:} For each voter in $Y$ that votes $\{a,b\} > \dots$ add a blue edge $(a,b)$.

\noindent
{\bf Uncolored edges:} For each candidate $c \in C-\{p\}$, add $b(c) = s_p$
uncolored edges $(c, x)$.

\smallskip

Suppose there exists a way to add $k$ voters from $X$ and $\ell$ voters from $Y$ such that $p$ wins.
As mentioned above, we can assume that we add as many voters approving $p$ as possible and so the score of $p$ is $s_p$.

We construct a matching in the graph $G$ by taking the $k$ red edges corresponding to the voters added from $X$ and the $\ell$ blue edges corresponding to the voters added from $Y$. Since the score of $p$ is $s_p$, vertex $p$ is incident to $s_p = b(p)$ edges in the matching. Since $p$ is a winner, every other candidate $c$ has score at most $s_p$, and so vertex $c$ is incident to at most $s_p = b(c)$ edges in the matching. What remains is to make this a perfect matching. 
The sum of the $b$-values for the vertices in $C$ is $\|C\|s_p$.
Since there are $k+\ell$ edges in the matching, $\|C\|s_p - 2(k+\ell)$ uncolored edges must be added to vertices in $C-\{p\}$ (since $k + \ell \geq s_p$, we have enough uncolored edges). We add exactly $b(x)$ uncolored edges and so this is a perfect matching.

For the converse, suppose there exists a perfect matching with $k$ red edges and $\ell$ blue edges. For each of the $k$ red edges add the corresponding voter from $X$, and for each of the $\ell$ blue edges add the corresponding voter from $Y$. Since this is a perfect $b$-matching, vertex $p$ is incident to exactly $s_p$ colored edges in the matching.
And each vertex $c \in C-\{p\}$ is incident to at most $s_p$ colored edges in the matching. It is easy to see that when the voters in $X$ are added that correspond to the red edges and the voters in $Y$ are added that correspond to the blue edges that $p$ wins.~\end{proofs}

What is left to show for equivalence is the reduction in the other direction.
The proof of the following theorem has been deferred to the appendix.

\begin{theorem}\label{thm:match-2-app-exact-ccrv}
\exactperfectmatching\ logspace many-one reduces to 2-Approval-CCRV!
\end{theorem}

\def\restrictedmatching{%

\begin{lemma}\label{lem:half-exact}
\exactperfectmatching\ logspace many-one reduces to Restricted Exact Perfect Matching.
\end{lemma}

\begin{proofs}
We denote the number of vertices of $G$ by $n$ and we will assume that $n$ is even. %
If $\|E'\| > n/2$, add enough isolated nonred edges to ensure that the number of red edges is exactly half the number of vertices. Specifically, add $\|E'\| - n/2$ new nonred isolated edges. Then the number of vertices is $n + 2\|E'\| - n$, and the number of red edges is exactly half of that. In addition, there is a perfect matching with exactly $\ell$ red edges in $G$ if and only if there is a perfect matching with exactly $\ell$ red edges in the padded graph.

If $\|E'\| < n/2$, Let $\hat{n}$ be the smallest even integer such that adding a red clique of size $\hat{n}$ will make the number of red edges at least half of the number of vertices. That is, $\hat{n}$ is the smallest even integer such that $\|E'\| + \hat{n}(\hat{n} - 1)/2 \geq (n + \hat{n})/2$. Then add $\hat{n}$ new vertices consisting of $\hat{n}/2$ red edges and enough other red edges to ensure that the number of red edges is exactly half of the number of vertices. Note that $G$ has a perfect matching containing exactly $\ell$ red edges if and only if the padded graph has a perfect matching containing exactly $\ell + \hat{n}/2$ red edges.~\end{proofs}
}

\subsection{2-Approval-CCRV is in P}\label{s:2app-ccrv}

In the previous section we showed that 2-Approval-CCRV! is logspace many-one equivalent to \exm.
However, unlike the case for First-Last, we did not get the same equivalence for CCRV. We only got the upper bound, showing that 2-Approval CCRV is in $\rp$, and thus easy. The obvious question is whether we can strengthen this result by showing equivalence with \exactperfectmatching\ or by showing that the problem is in $\p$. In this section, we will show the latter.

Note that the lower bound proof for First-Last-CCRV depended on the lower bound for First-Last-CCAV!.
However, 2-Approval-CCAV! is in $\p$ (see appendix).

\begin{theorem}\label{thm:2app-ccav-exact}
2-Approval-CCAV! is in \p.
\end{theorem}

We will now show that 2-Approval-CCRV is also in $\p$.

As a first attempt to obtain a better upper bound for 2-Approval-CCRV, it makes sense to look at the logspace many-one reduction from 2-Approval-CCRV! to \exactrbbm\ (Theorem~\ref{thm:2app-ccrv-exact-upper}). For this attempt we will define an appropriate modification of \exactrbbm, so that we can modify this reduction into a reduction from 2-Approval-CCRV to the new variation of \exactrbbm.

Recall that we viewed CCRV! as a version of exact control by adding voters
where we have two sets of unregistered voters $X$ and $Y$ and we ask if it is possible to add (to the empty set)
$k = \|X\|-\ell$ voters from $X$ and $\ell$ voters from $Y$ such that $p$ wins. 
This then suggested reducing 2-Approval-CCRV! to \exactrbbm.

We can similarly view CCRV as a version of control by adding voters
where we have two sets of unregistered voters $X$ and $Y$ and we ask if it is possible to add (to the empty set)
$k' \geq \|X\|-\ell$ voters from $X$ and $\ell' = \|X\| - k'$ voters from $Y$ such that $p$ wins. (Note that $k' + \ell' = \|X\|$.)

This suggests the following modification of \exactrbbm.

\nonameprob%
{A multigraph $G = (V,E)$, a capacity function $b: V \rightarrow \naturalnumber$, disjoint sets $R \subseteq E$ of red edges and $B \subseteq E$ of blue edges, an integer $k \geq 0$, and an integer $\ell \geq 0$.}
{Does there exist a perfect $b$-matching of $G$ with exactly $k' \geq k$ red edges and exactly $\ell' = \ell - (k' - k)$ blue edges, i.e., a set of edges $E' \subseteq E$, exactly $k + \ell$ of which are colored and at least $k$ of which are red, such that each $v \in V$ is incident with exactly $b(v)$ edges in $E'$?}

It is not hard to show that the proof of Theorem~\ref{thm:2app-ccrv-exact-upper} can be minorly modified to establish that 2-Approval-CCRV indeed logspace many-one reduces to this modification of \exactrbbm.
However, \exm\ reduces to this modification of \exactrbbm,
since we need to have an exact number of colored edges in the matching. So this will not show that 2-Approval-CCRV is in $\p$.

To show 2-Approval-CCRV in $\p$, it turns out that it makes sense to look at the proof that 2-Approval-CCAV! is in $\p$~(Theorem~\ref{thm:2app-ccav-exact}). We showed this by
reducing 2-Approval-CCAV! to Max-Cardinality $b$-Matching for Multigraphs.
We will reduce
2-Approval-CCRV to Max-Weight $b$-Matching for Multigraphs, which is defined as follows.

\prob{Max-Weight $b$-Matching for Multigraphs}%
{An edge-weighted multigraph $G = (V,E)$,
a function $b: V \to \mathbb{N}$, and
integer $k \geq 0$.}%
{Does there exist an $E' \subseteq E$ of weight at least $k$ such
that each vertex $v \in V$ is incident to at
most $b(v)$ edges in $E'$?}

Max-Weight $b$-Matching for Multigraphs is in P~\cite{edm-joh:c:matching}.\footnote{As explained in~\citeauthor{ger:b:matching}~(\citeyear{ger:b:matching}, Section 7), it is easy to reduce such problems to 
Max-Weight Matching, which is well-known to be in P~\cite{edm:j:matching}, using the construction from~\cite{tut:j:factor}. (Note that we can assume that the $b$-values are bound by the number of edges in the graph.)
For the specific case of Max-Weight $b$-Matching for Multigraphs, we can modify the reduction described in the proof of Theorem~\ref{t:XbtoX} as follows: Edges corresponding to original edges keep their weight. Padding edges get weight $W$, where $W$ is a positive integer greater than the sum of all the edge weights in the original graph. This ensures that the number of padding edges will be maximized in a max-weight matching.}
Though many matching problems are in $\p$ (see, e.g.,~\citeauthor{ger:b:matching}~(\citeyear{ger:b:matching}, Section 7)), minor-seeming variations can make these problems hard. As mentioned earlier in this paper \exm\ %
is in $\rp$ but not known to be in $\p$. It is worth mentioning that the weighted version is $\np$-complete, since Subset Sum straightforwardly reduces to it~\cite{pap-sch-yan:local-search}. In the context of voting problems, the weighted version of CCAV for 2-Approval
can be viewed as a variation of a weighted matching problem~\cite{lin:thesis:elections}. Since weighted CCAV for 2-Approval is $\np$-complete~\cite{fal-hem-hem:j:weighted-control}, so is the corresponding matching problem.

To show the main result of this section we adapt
our proof of Theorem~\ref{thm:2app-ccrv-exact-upper} (2-Approval-CCRV! logspace many-one reduces to \exactrbbm).

\begin{theorem}\label{thm:2app-ccrv-upper}
2-Approval-CCRV logspace many-one reduces to Max-Weight $b$-Matching for Multigraphs.
\end{theorem}

\begin{proofs}
Consider an instance of 2-Approval-CCRV with
candidates $C$, registered voters $X$, unregistered voters $Y$, preferred candidate $p$, and number $\ell$.
Similarly to the proof of Theorem~\ref{thm:2app-ccrv-exact-upper}, we
view this problem as a version of control by adding
voters where we now ask if it is possible to add (to the empty set)
$k' \geq \|X\|-\ell$ voters from $X$ and $\ell' = \|X\| - k'$ voters from $Y$ such that $p$ wins.

Note that we can assume that we will
add as many voters that approve of $p$ as possible. This fixes the final score of $p$ which we denote by
$s_p$. Let $X_p$ be the set of voters in $X$ approving $p$. Then 
$s_p = \|X_p\| + \min(\ell, \|X\| - \|X_p\|, \text{\#voters in $Y$ approving $p$})$. 

We will now construct the instance of Max-Weight $b$-Matching for Multigraphs.

Let the set of vertices in the constructed graph $V(G) = C \cup \{x\}$, where $x$ is a vertex used to ensure that the resulting matching contains exactly $\|X\| = k' + \ell'$ edges between elements of $C$ (which will correspond to the set of added voters).
For each $c \in C$ let $b(c) = s_{p}$. %
Let $b(x) = \|C\|s_p - 2\|X\|$.
If this is negative, control is not possible.
The set of edges is defined as follows.

\begin{description}
    \item[Heavy edges] For each candidate $c \in C-\{p\}$, add $b(c) = s_p$ edges $(c, x)$ of weight $H$, where $H$ is larger than the sum of all light edges combined. This implies that any max-weight matching will maximize the number of heavy edges, and so any max-weight matching will contain $b(x) = \|C\|s_p - 2\|X\|$ heavy edges and at most $\|X\|$ light edges.
    \item[Light edges corresponding to $X$] For each voter in $X$ that votes $\{a,b\} > \dots$ add an edge $(a,b)$ of weight $\|X\| + 1$.
    \item[Light edges corresponding to $Y$] For each voter in $Y$ that votes $\{a,b\} > \dots$ add an edge $(a,b)$ of weight $\|X\|$.
\end{description}

We will show that $p$ can be made a winner by replacing $\leq \ell$ voters if and only if $G$ has a max-weight matching of size $\geq b(x)H + \|X\|^2 + (\|X\| - \ell)$.

Suppose $p$ can be made a winner by replacing at most $\ell$ voters. As mentioned above, we can assume that we add as many voters approving $p$ as possible. Suppose we add $k' \geq \|X\|-\ell$ voters from $X$ and $\ell' = \|X\| - k'$ voters from $Y$ such that $p$ wins. The score of $p$ is $s_p$.

We construct a matching in the graph $G$ by taking the $k'$ edges corresponding to the voters added from $X$ and the $\ell'$ edges corresponding to the voters added from $Y$. Since the score of $p$ is $s_p$, vertex $p$ is incident to $s_p = b(p)$ edges in the matching. Since $p$ is a winner, every other candidate $c$ has score at most $s_p$, and so vertex $c$ is incident to at most $s_p = b(c)$ edges in the matching. 
The sum of the $b$-values of the vertices in $C$ is $\|C\|s_p$. 
Since there are $\|X\|$ light edges in the matching, $\|C\| s_p - 2 \|X\| = b(x)$ heavy edges can be added to the matching.
Note that the weight of this matching is $b(x)H + \|X\|^2 + k' \geq b(x) H + \|X\|^2 + (\|X\| - \ell)$.

For the converse, suppose there exists a matching with weight $\geq b(x)H + \|X\|^2 + \|X\| - \ell$. Then this matching contains exactly $b(x)$ heavy edges and at most $\|X\|$ light edges. If there were fewer than $\|X\|$ light edges in the matching, then the weight of the light edges would be at most $(\|X\| - 1)(\|X\| + 1) < \|X\|^2$. It follows that the matching contains exactly $\|X\|$ light edges and $\geq \|X\| - \ell$ of these correspond to voters in $X$. Since this is a perfect $b$-matching, vertex $p$ is incident to exactly $s_p$ light edges in the matching
and each vertex $c \in C-\{p\}$ is incident to at most $s_p$ light edges in the matching. It is easy to see that when the $\|X\|$ voters corresponding to light edges in the matching are added, $p$ is a winner. And at least $\|X\| - \ell$ of these light edges correspond to voters in $X$~\end{proofs}

Since Max-Weight $b$-Matching for Multigraphs
is in \p, the above theorem immediately implies the  main result of this section.

\begin{corollary}\label{cor:2-app-ccrv}
2-Approval-CCRV is in \p.
\end{corollary}

\section{Conclusion}

We showed that by considering \rp\ as an option, we gain insight into the complexity of voting problems. We connected the complexity of
control by replacing voters for First-Last and 2-Approval elections to the complexity of the \exbioptm, showing these problems in
\rp. We expect this approach will be useful in exploring
the complexity of other voting problems.
For 2-Approval control by replacing voters this ultimately led to showing the problem to be in \p, which solves the last remaining case in the comprehensive table from Erd{\'e}lyi et al.~\shortcite{erd-nev-reg-rot-yan-zor:j:towards-completing}.

\section*{Acknowledgements}

This work was supported in part by NSF-DUE-1819546.
We thank Lane Hemaspaandra for helpful discussions and the anonymous reviewers for helpful comments and suggestions.

\appendix

\section{Appendix}

\noindent
{\bf Theorem~\ref{t:XbtoX}}\ {\em \exbioptbm\ logspace many-one reduces to \exbioptm.}

\smallskip

\XbToXproof

\noindent
{\bf Theorem~\ref{t:fl-ccrv-to-exexbbm}}\ {\em First-Last-CCRV! logspace many-one reduces to \exexbibm.}

\smallskip

\begin{proofs}
Let $X$ be the set of registered voters, $Y$ be the set of unregistered voters, and $z$ the number of voters to replace.
As mentioned previously, we view this problem as adding $\|X\| - z$ voters from $X$ and $z$ voters from $Y$. We preprocess as in the proof of Theorem~\ref{t:CCAVtoM} to find out (in logspace) whether control is possible or to construct an instance of exact control by adding voters with two sets of unregistered voters that each give 0 points to $p$. Let $V$ be the set of registered voters, $W_1$ and $W_2$ be the sets of unregistered voters, and $k$ and $\ell$ be the bounds. And we are asking if we can add $k$ voters from $W_1$ and $\ell$ voters from $W_2$ such that $p$ is a winner.

We now proceed as in the construction from Theorem~\ref{t:CCAVtoM}, with the obvious adaptation for the coloring of the edges: Edges corresponding to voters in $W_1$ are colored red, edges corresponding to voters in $W_2$
are colored blue, and the remaining edges are uncolored.

The same argument as in the correctness of the construction of Theorem~\ref{t:CCAVtoM} shows that we can add $k$ unregistered voters from $W_1$ and $\ell$ unregistered voters from $W_2$ such that $p$ becomes a winner if and only if there exists a perfect matching that contains exactly $k$ red edges and exactly $\ell$ blue edges.~\end{proofs}

\noindent
{\bf Theorem~\ref{thm:2app-ccav-exact}}\
{\em 2-Approval-CCAV! is in \p.}

\smallskip

\begin{proofs}
We show 2-Approval-CCAV! in \p\ by a straightforward reduction to Max-Cardinality $b$-Matching for Multigraphs, which is in P~\cite{edm-joh:c:matching}. The proof is similar to the proof that 3-Approval-CCAV and 2-Approval-CCDV are in \p~\cite{lin:thesis:elections}.
\prob{Max-Cardinality $b$-Matching for Multigraphs\footnote{This problem is called Simple $b$-Edge Matching for Multigraphs in Lin~\shortcite{lin:thesis:elections}.}}%
{A multigraph $G = (V,E)$,
a function $b: V \to \mathbb{N}$, and
integer $k \geq 0$.}%
{Does there exist an $E' \subseteq E$ such
that $\|E'\| = k$ and each vertex $v \in V$ is incident to at
most $b(v)$ edges in $E'$?}

Let $k$ be the number of unregistered voters to add. If there are at least $k$ unregistered voters voting for $p$, then it is clearly in \p\ to determine which $k$ voters to add. Otherwise, add all unregistered voters voting for $p$. Redefine $k$ to be the number of unregistered voters, who now all do not vote for $p$, to add. Let $s_c$ be the current score of candidate $c$.
For each $a \neq p$, let $b(a)$ the maximum number of points that we can add to $a$, i.e., $s_p - s_a$. If any of the $b(a)$'s is negative, control is not possible. Otherwise, we turn this into an
instance of Max-Cardinality $b$-Matching for Multigraphs. For each unregistered vote $\{a,b\} > \cdots$
add an edge between $a$ and $b$. And we ask if there is a $b$-matching of size at least $k$.~\end{proofs}

\noindent
{\bf Theorem~\ref{thm:match-2-app-exact-ccrv}}\ {\em \exactperfectmatching\ logspace many-one reduces to 2-Approval-CCRV!}

\smallskip

\begin{proofs}
We mention that a similar reduction is used to reduce \exactperfectmatching\ to a different matching problem called Blue-Red Matching~\cite{DBLP:conf/mfcs/NomikosPZ07}, though this problem is not know to be equivalent to \exactperfectmatching.

We reduce from a variant of \exactperfectmatching\ defined below where the number of red edges is exactly half the number of vertices.

\prob{Restricted Exact Perfect Matching}%
{A graph $G = (V,E)$, a set of red edges $E' \subseteq E$ of size $\|V\|/2$ and an integer $\ell \geq 0$}%
{Does $G$ contain a perfect matching that contains exactly $\ell$ edges from $E'$?}

We show in Lemma~\ref{lem:half-exact}
that \exactperfectmatching\ reduces
to Restricted Exact Perfect Matching. %

We now reduce Restricted Exact Perfect Matching to
2-Approval-CCRV!.
Given a graph $G = (V,E)$ with $n$ vertices, a set of red edges $E' \subseteq E$ with $\|E'\| = n/2$, and an integer $\ell \geq 0$,
we construct the instance of 2-Approval-CCRV! as follows.

Let the set of candidates $C = V(G) \cup \{p,p'\}$, and let the preferred candidate be $p$. For each edge $(a,b)$ in $G$, we have a voter voting $\{a,b\} > \cdots$. If the edge is colored red, this voter is a registered voter. If the edge is not red, this is an unregistered voter. Add one additional registered voter voting $\{p,p'\} > \cdots$, and so each candidate $C-\{p\}$ can score at most 1 in an instance where $p$ is a winner.
Let the number of voters to replace be $n/2 - \ell$.

Suppose there exists a perfect matching that contains exactly $\ell$ red edges (and $n/2 - \ell$ nonred edges). It is easy to see that control is possible by replacing the $n/2 - \ell$ registered voters corresponding to the red edges not in the matching with the $n/2 - \ell$ unregistered voters
corresponding to the nonred edges in the matching. Since this is a perfect matching, each vertex is incident to exactly one edge in the matching. It is easy to see that every candidate has score 1, and so $p$ is a winner.

For the converse, suppose there exists a way to replace $n/2-\ell$ registered voters with $n/2-\ell$ unregistered voters such that $p$ wins. $p$ is approved by just one voter and it is easy to see that this voter should not be replaced. Since $p$ is a winner the score of each candidate $c \in C-\{p\}$ is at most 1.
Construct a matching as follows. For each of the registered voters not replaced, add its
corresponding red edge to the matching. For each of the unregistered voters added to the election, add its corresponding nonred edge. Since the number of voters replaced is $n/2-\ell$, this is a perfect matching with $\ell$ red edges.~\end{proofs}
\label{app:restricted-matching}
\restrictedmatching

\end{document}